\documentstyle[12pt]{article}
\input tcilatex

\begin{document}

\title{Anderson's localization in a random metric: applications to cosmology}
\author{J. C. Flores and M. Bologna}
\date{Departamento de F\'{i}sica, Universidad de Tarapac\'{a}, Casilla 7-D, Arica,
Chile}
\maketitle

\baselineskip=14pt

\bigskip

{\bf Abstract:} It is considered an equation for the Lyapunov exponent $%
\gamma $ in a random metric for a scalar propagating wave field. At first
order in frequency this equation is solved explicitly. The localization
length $L_{c}$ (reciprocal of Re($\gamma $)) is obtained as function of the
metric-fluctuation-distance $\Delta R$ (function of disorder) and the
frequency $\omega $ of the wave. Explicitly, low-frequencies propagate
longer than high, that is $L_{c}\omega ^{2}=C^{te}$. Direct applications
with cosmological quantities like background radiation microwave ($\lambda
\sim 1/2\times 10^{-3}$ [m]) and the Universe-length ( `localization length' 
$L_{c}\sim 1.6\times 10^{25}$ [m]) permits to evaluate the
metric-fluctuations-distance as $\Delta R\sim 10^{-35}$ [m], a number at
order of the Planck's length.

\[
\]


72.15.Rn (Anderson's localization)

98.80.-k (Cosmology)

04.40.-b (Classical fields in general relativity)

\newpage

{\Large I Introduction: Anderson's localization.}

\[
\]

Anderson's localization [1-3] is a general non dissipative phenomena related
to wave localization due to interference in random (disordered) media.
Depending on the degree of disorder ($\sigma $), frequency ($\omega $) and
dimension of the space ($D$) waves could remain exponentially localized in
some region characterized by the so-called localization length ($L_{c}$).
Certainly, the calculation of the localization length and the spectrum for
these systems is not an easy task, nevertheless, some general properties are
today well understood [2,3]. It is known that Anderson's localization is
very sensitive to correlation between different random parts of the media
[4-9] and decoherence [10,11]; it has been studied in systems like seismic
waves (coda) [12,14], electromagnetic [15] and sound waves [16], electronic
systems [17], chaos [18] and others. Anderson's localization is a general
phenomena related to waves and it seems natural to investigate its
application on waves propagating in a given random metric, explicitly in a
general relativity context as, for instance, waves in the whole space
(Universe). Indeed being the curvature of the space-time described by the
metric tensor, that is directly related to the matter distribution, it is
reasonable to assume that, when the matter is distributed in a random way,
the metric also can be considered random. In this case wave localization
properties in these random systems could be expected.

In this paper we will found an expression for the localization length $L_{c}$
for waves propagating in a random metric (ground $\oplus $ disorder). The
metric will be characterized and the basic parameter related to it, the
metric-fluctuations-distance $\Delta R$ could be calculated from the ground
metric and disorder (both assumed translation invariant). In the limit of
slow frequency, the localization length $L_{c}$ as function of the
metric-fluctuations will be obtained and given by (section III. $c$ light
speed)

\begin{equation}
L_{c}=\frac{c^{2}}{\omega ^{2}\Delta R}.
\end{equation}
An expression quite reasonable since $L_{c}=\infty $ when $\Delta R=0$ (no
disorder) or when $\omega =0$. In fact, is well known that for classical
waves with long wavelength are less localized since it does not `feel'
inhomogeneities (disorder).

It seems quite intriguing that when we consider the background microwave
radiation ($\lambda \sim 1/2\times 10^{-3}$[m]) and identifying the universe
radii with the localization length ($L_{c}\sim 1.6\times 10^{25}$ [m]) the
above expression (1) gives a metric-fluctuation-distance similar to the
Planck's length ($\Delta R\sim 10^{-35}$ [m], section IV).

In the next section (II), the basic equation (Riccati) defining the Lyapunov
exponent \ (vector) in a random metric will be presented. A formal
frequency-expansion will be proposed for the real and imaginary part of the
Lyapunov exponent. The real part is identified with the reciprocal of the
localization length $L_{c}$. In section III the equation is solved formally
for uncorrelated disorder in the small frequency limit. The assumption of
independent random variables for the temporal and spatial part of the metric
is always assumed. The localization length is expressed in term of Green
functions. Nevertheless, defining the metric-fluctuation-distance, the
localization length \ could be expressed in a more intuitive way (equation
(1) or (21)). As pointed out in this introduction, in section IV we will
study a direct application using cosmological parameters. We find that
disorder fluctuations are of order of the Planck's length, a quite
intriguing result.

\[
\]

{\Large II Lyapunov exponent.}

\[
\]

Let us consider a scalar field $\phi (x_{\mu })$ defined by a wave equation
in a four dimensional time independent symmetric metric $g^{\alpha \beta }$,
\ where we assume $g^{01}=g^{02}=g^{03}=0$. Explicitly, consider the
wave-equation [19,20] for the scalar field:

\begin{equation}
-\omega ^{2}g^{oo}\phi +g^{ik}\phi _{;i;k}=0,\,\,\,\,i,k=1,2,3.
\end{equation}
where $g^{oo}>0$ and the symbol $;i$ define the covariant derivative respect
to the spatial coordinate $x_{i}$. Latin indices like $i,j,$ etc., run over
three spatial coordinate. As usual, the controvariant tensors are related to
covariant ones by the metric tensor. We will assume that the field $\phi $
does not affect the metric, namely, $g\neq g(\phi )$ as expected for weak
propagating fields. Also note that we use units where the light velocity $%
c=1 $.

>From a general point of view, we consider a random tensor metric like to $g=$
$g_{(o)}+\Delta g$ where $g_{(o)}$ denotes the translation-invariant
background deterministic metric and $\Delta g$ a random tensor which will be
specified later. Since we are interested in Anderson's localization
phenomena it is useful to consider the (local) Lyapunov exponent $\gamma
_{i},$ which is a vector, defined by

\begin{equation}
\gamma _{i}=\frac{\phi _{;i}}{\phi }R,
\end{equation}
where $R$ is a parameter with dimension of spatial distance and assumed an
invariant of the system, it will be specified later. In fact, it does not
play an important role and we use it only for dimensional reasons. The
reciprocal of the norm of real part of $\gamma _{i}$ will be related with
the localization length. From (2), the equation for the components of $%
\gamma $ becomes

\begin{equation}
R\gamma _{;i}^{i}=R^{2}\omega ^{2}g^{00}-\gamma ^{i}\gamma _{i}.
\end{equation}
We stress that the covariant derivative in (4) is respect to the spatial
metric (i.e. $g^{ij}$ ). \ Eq.(4) is the start-point to our calculations and
corresponds to a Riccati type equation.

The real and the imaginary part of the Lyapunov exponent are defined as $%
a_{i}=Re\,\gamma _{i}$ and $b_{i}=Im\, \gamma _{i}$. Namely,

\begin{equation}
\gamma _{i}=a_{i}+\sqrt{-1}b_{i}.
\end{equation}
and the norm of the real part of $\gamma _{i}$ will be identified with the
inverse of the localization length and the imaginary part with the
wave-vector. From the equation (4) for the Lyapunov exponent, we have the
pair of coupled equations for these components

\begin{equation}
Ra_{;i}^{i}=R^{2}\omega ^{2}g^{00}-a^{i}a_{i}+b^{i}b_{i},
\end{equation}
and

\begin{equation}
Rb_{;i}^{i}=-2a^{i}b_{i},
\end{equation}
which are exact. Nevertheless, taking in account that the metric is random,
for any realization we must use the corresponding covariant derivative. For
the euclidean metric the wave-plane solution of (6,7) is $a_{i}=0$ and $%
b_{i} $=constant, with the restriction $R^{2}\omega ^{2}+b^{i}b_{i}=0$ on
the norm of $b$.

It is worthy to stress that for the above pair of equations, the solution $%
a=b=0$ corresponds to the frequency $\omega =0$ and is formally an extended
state ($\phi =$ constant) so that we could expect for low frequency a large
localization length. This fact is well known for classical waves but not
necessarily true for electronics waves where disorder does not multiplies
the frequency. In fact this limit (slow frequency) can be treated
analytically and it will be the object of the next section. We consider the
formal expansion on frequency:

\begin{equation}
a_{i}=\sum_{n=1}^{\infty }\left( R\omega \right) ^{2n}A_{i}^{(n)},
\end{equation}
and

\begin{equation}
b_{i}=\sum_{n=1}^{\infty }\left( R\omega \right) ^{2n-1}B_{i}^{(n)},
\end{equation}
where the news amplitudes $A$ and $B$ must be determined by solving (6,7).
The choice $\omega $-even for the function $a$ and $\omega $-odd for $b$
becomes directly from the form of these equation. We do not write explicitly
the evolution equation for all order in the amplitude $A$ or $B$ because is
not relevant for our purposes. In the next section we will consider the
first order only.

\[
\]

{\Large III \ \ Small frequency case: calculation of the localization length.%
}

\[
\]

As said before, in this section we will consider the small frequency
expansion and the localization length will be evaluate. Consider the first
order expansion of equations (8,9) for the Lyapunov exponent given by
(section II) 
\begin{equation}
a_{i}=R^{2}\omega ^{2}A_{i}+O(\omega ^{4}),\,\,\text{and}\,\,b_{i}=R\omega
B_{i}+O(\omega ^{3}).
\end{equation}
Where $A$ and $B$ are auxiliary variables. From (10) we have the pair of
equations: 
\begin{equation}
RA_{;i}^{i}=(g^{00}-1)+(1+B^{i}B_{i}),\,\,\text{and }\,\,B_{;i}^{i}=0.
\end{equation}
The amplitude $B$ will be the choose as a constant with $B^{i}B_{i}+1=0$. In
fact, this amplitude is related to the wave vector for homogenous systems.
On the other hand, the equation for $A$ becomes

\begin{equation}
RA_{;i}^{i}=\varepsilon (x),\text{ \,where\, }g^{00}=1+\varepsilon (x)\text{
\,and\, }x\equiv\left(x_1,x_2,x_3\right).
\end{equation}
The random independent variables $\varepsilon (x)$ (the disorder) are
assumed well characterized and depending only of spatial coordinates.
Explicitly, we assume:

\begin{equation}
\left\langle \varepsilon (x)\right\rangle =0\text{ and }\left\langle
\varepsilon (x)\varepsilon (x^{\prime })\right\rangle =R^{D}\sigma
^{2}\delta (x -x ^{\prime }),
\end{equation}
where the symbol $\left\langle \cdot \right\rangle $ denotes disorder
average, $\sigma $ is the dimensionless disorder parameter (dispersion) and $%
\delta $ is the usual Dirac distribution in the corresponding spatial
dimension $D,$ actually $D=3$. The above equations define our disordered
model for the temporal part of the metric.

The formal solution of (12) is given by 
\begin{equation}
A^{i}(x)=\frac{1}{R}\int dV^{\prime }G^{i}(x,x^{\prime })\varepsilon
(x^{\prime }),
\end{equation}
where $dV$ is an invariant element of volume and the spatial Green function $%
G$ is solution of $G_{;i}^{i}(x,x^{\prime })=\delta (x-x^{\prime })$. Note
that it depends on the random spatial metric which will not be explicitly
considered. In fact, Gauss theorem ensures that $G^{i}G_{i}=1/S^{2}$ where $%
S $ is the surface of the `sphere' of radius $r$ in the corresponding
metric. Taking the norm of this vector:

\begin{equation}
\Vert A^{i}(x)\Vert ^{2}\equiv -A^{i}(x)A_{i}(x)=-\frac{1}{R}\int dV^{\prime
}dV^{\prime \prime }G^{i}(x,x^{\prime })G_{i}(x,x^{\prime \prime
})\varepsilon (x^{\prime })\varepsilon (x^{\prime \prime }),
\label{normlyap}
\end{equation}
at this stage this scalar quantity is depending on the spatial variables.
Since we assume that the random components of the metric tensor (spatial and
temporal) are independents, using the Eq. (13), the ensemble average of the
norm of $A$ becomes 
\begin{equation}
\left\langle A^{i}A_{i}\right\rangle =\sigma ^{2}R^{D-2}\int \left\langle
dVG^{i}G_{i}\right\rangle ,
\end{equation}
which is a space invariant quantity since we assume homogenous disorder and
\ translation invariant background metric, so it is not depending on a
particular positions. \ Note that when $\left\langle g^{00}\right\rangle
\neq 1$ the above result is valid for the dispersion $\left\langle \left(
A^{i}A_{i}-\left\langle A^{i}\right\rangle \left\langle A_{i}\right\rangle
\right) \right\rangle $.

The norm of the real part for the Lyapunov exponent satisfies

\begin{equation}
\left\langle a^{i}a_{i}\right\rangle =\omega ^{4}\sigma ^{2}R^{D+2}\int
\left\langle dVG^{i}G_{i}\right\rangle ,
\end{equation}
and defining formally the localization length $L_{c}$ \ as

\begin{equation}
L_{c}^{-2}\equiv -\frac{1}{R^{2}}\left\langle a^{i}a_{i}\right\rangle ,
\end{equation}
the following expression is obtained:

\begin{equation}
L_{c}^{-2}=-\omega ^{4}\sigma ^{2}R^{D}\int \left\langle
dVG^{i}G_{i}\right\rangle .
\end{equation}
It is relevant to note that:

\[
\]

(i) The integral is only on spatial coordinates and we need more
specification to do it. Moreover, it depends on the dimension $D$.

(ii) Low frequencies have large localization length as expected for
classical waves in random media. In fact, $L_{c}^{-2}\omega ^{4}=$constant,
for a given random system.

(iii) The localization length is a global property of the systems like to
the averaged Lyapunov exponent (no position depending).

(iv) The invariant distance $R$ is assumed, for instance, as related to the
inverse mean density $\rho $ of matter (i.e. $R\sim \rho ^{-1/D}$).
Nevertheless, eventually, it could be also considered other relevant
distance parameter of the systems.

\[
\]

Defining the disorder parameter $\Delta R$ (the
metric-fluctuations-distance) as 
\begin{equation}
\left( \Delta R\right) ^{2}\equiv -\sigma ^{2}R^{D}\int \left\langle
dVG^{i}G_{i}\right\rangle ,
\end{equation}
the localization length becomes in term of this parameter given by

\begin{equation}
L_{c}^{-2}=\omega ^{4}\Delta R^{2}/c^{4},
\end{equation}
where we have restored the velocity of light $c$. Note that the
metric-fluctuations-distance $\Delta R$ is a parameter which must be
calculated using the distribution of spatial disorder. Expression (1) \ in
the section I corresponds essentially to (21).

\[
\]

{\Large IV \ \ The microwave radiation background, localization length, and
metric fluctuation estimations: Planck's length. }

\[
\]

As application of the above results, consider scalar waves propagating in
the whole Universe assumed as a media with disordered-metric. We will
consider explicitly microwave background radiation and the radii of the
Universe identified as its associated localization length. It must be noted
here that the background radiation is a electromagnetic vector field, not a
scalar-field. Nevertheless, every component of this field can be formally
reduced to a scalar evolution equation. We assume \ that the metric is
slowly changing in time \ compared to others time scale.

The expression for the localization length (21) could be written in term of
the wavelength $\lambda $ \ as ($L_{c}=\left( L_{c}^{-2}\right) ^{-1/2}$)

\begin{equation}
L_{c}=\frac{\lambda ^{2}}{\left( 2\pi \right) ^{2}\Delta R}.
\end{equation}

At this point a remark becomes in order, for disordered systems the wave
vector is not a parameter characterising the wave since the whole system is
not translation invariat. \ So, the definition of the wavelength is in the
sence of $\lambda =2\pi c/\omega $. For instance, assuming a wavelength $%
\lambda \sim \frac{1}{2}\times 10^{-3}$[m] for the microwave radiation
background at the universe, and $L_{c}\sim 1.6\times 10^{25}$ [m] [20] for
the universe size then, we obtain for the metric fluctuations $\Delta R\sim
3.9\times 10^{-35}$ [m]. Namely, fluctuations in the metric in order of the
Planck's length. It is a surprising result since no direct relation-ship a
prior exist between Anderson's localization and Planck's length. Note that
the condition of small frequencies hold since the metric fluctuations are
smaller than the wave length.

\[
\]

{\bf Conclusion:} Anderson's localization was directly considered from a
Riccati type equation (4) with a random metric with separated spatial and
temporal disorder (uncorrelated). The localization length in the
low-frequency limit was evaluated (eq.(1) or (21)). It is a quit reasonable
expression, that is, for low-frequency the localization length goes to
infinite and also for small disorder. A direct application to the background
microwave radiation at the Universe, identifying the universe length with
the localization length, gives us an estimation of the
metric-length-fluctuations. It corresponds to the Planck's length. A
surprising result since Planck's length is not put in the calculation as
input.

\[
\]

{\bf Acknowledges}. JCF was supported by Grant FONDECYT \ 1040311 and \
acknowledges interesting conversations carried-out with D. Alloin (ESO).
Useful sugestions were give to us by L.Tessieri and T. Zannias (Universidad
de Michoacana). Also, we express our acknowledge to C. Leiva (UTA) for
valuable comments.

\end{document}